\documentclass[12pt,preprint]{aastex}

\usepackage{graphicx}

\shorttitle{Distinct population in the KDC of NGC~1700}
\shortauthors{Kleineberg et  al.}

\begin{document}

%



\title{Evidence of a distinct stellar population in the counter-rotating core of NGC~1700}
\author{K. Kleineberg}
\affil{Instituto de Astrof\'{\i}sica de Canarias, E38200, La Laguna, Tenerife, Spain\\
       Departamento de Astrof\'{\i}sica, Universidad de La Laguna, E38205, la Laguna, Tenerife, Spain}
\author{P. S\'anchez-Bl\'azquez}
\affil{Departamento de F\'{\i}sica Te\'orica, Universidad Aut\'onoma de Madrid, Cantoblanco, E28049, Madrid, Spain}
\author{A. Vazdekis}
\affil{Instituto de Astrof\'{\i}sica de Canarias, E38200, La Laguna, Tenerife, Spain\\
       Departamento de Astrof\'{\i}sica, Universidad de La Laguna, E38205, la Laguna, Tenerife, Spain}
\altaffiltext{1}{Instituto de Astrof\'{\i}sica de Canarias, E38200, La Laguna, Tenerife, Spain}
\altaffiltext{2}{Departamento de Astrof\'{\i}sica, Universidad de La Laguna, E38205, la Laguna, Tenerife, Spain}

\begin{abstract}
We find a distinct stellar population in the counter-rotating and kinematically
decoupled core of the isolated massive elliptical galaxy NGC~1700. Coinciding
with the edge of this core we find a significant change in the slope of the
gradient of various representative absorption line indices. Our age estimate for
this core is markedly younger than the main body of the galaxy. We find 
lower values for the age, metallicity and Mg/Fe abundance ratio in the center 
of this galaxy when we compare them with other isolated elliptical galaxies with similar velocity
dispersion.  We discuss the different possible scenarios that might have lead to
the formation of this younger kinematically decoupled structure and conclude that,
in light of our findings, the ingestion of a small
stellar companion on a retrograde orbit is the most favoured.
\end{abstract}


\section{Introduction}

Galaxies with kinematically decoupled cores (KDC, hereafter) are very
interesting systems to test ideas of galaxy formation. The SAURON survey found
that $\sim$30 per cent of early-type galaxies have KDCs (de Zeeuw et al. 2002;
McDermid et al. 2006). Their distinct angular momenta suggest that the cores may
be relics of interactions or mergers and thus may provide a diagnostic
to estimate the relevance of these processes in shaping galactic properties
(e.g., Kormendy 1984; Franx \&
Illingworth 1988; Schweizer et al.\ 1990; Bender \& Surma 1992).

The origin of the KDC is still a mystery (Illingworth \& Franx 1989; Bender 1990), 
but the coupling between kinematic and stellar population studies can  provide
the necessary clues for achieving a more complete picture.
Several possible scenarios have been proposed (see Bender 1990 and references therein):  
(1) geometrical projection of the stellar orbits in the core of an 
ordinary triaxial potential (Statler 1991). Obviously,  
breaks in the 
stellar population properties of the KDC and the rest of the 
galaxy are not expected in this case;
(2) a central stellar disc formed from accreted gas with an initial
counter-rotating orbit (Franx \& Illingworth 1988; Bertola et al. 1998). Under
this process, however, it would not be possible to form some of the observed massive KDC. For
example,  the mass of the KDC of NGC~5322 is $\sim$10$^{10}$ M$_{\odot}$ (Bender
\& Surma 1992), which is significantly higher than the gaseous mass that an
irregular gaseous galaxy would contribute with; 
(3) a central disc formed during a major merger of two spirals (Schweizer et
al.\ 1990; Hernquist \& Barnes 1991). The gas and stars react very differently to the merger process and
the angular momentum of both components can be very different, leading to this
kind of structures,  and 
(4) the remnant of an accreted stellar companion, initially orbiting contrary to
the mean rotation of the galaxy (Kormendy 1984; Balcells \& Quinn 1990). The
companion would be partially destroyed but its nucleus would become part of the
decoupled core of the host galaxy. In this scenario we expect to find
breaks between the stellar populations of the KDC and the rest of the
galaxy. Low-mass galaxies have, in general, younger populations and lower
metallicities than their massive counterparts (e.g., S\'anchez-Bl\'azquez et al.
2006b), so we would expect to find a drop in these two parameters; 

So far, a clear difference in the stellar populations of
KDC and the host galaxy has not been found,  at least in galaxies with
central velocity dispersion above  100 kms$^{-1}$ 
(eg.,  Carollo et al. 1997; Mehlert et al. 1998; Davies
et al. 2001; S\'anchez-Bl\'azquez 2004; Kuntschner et al. 2010). Several authors have reported that KDCs
show enhanced metallicity using colours (eg., Bender \& Surma 1992; Carollo
\& Danzinger 1994b). However,
S\'anchez-Bl\'azquez (2004) shows that these variations were more likely due to the
presence of dust discs by comparing the measurements in colours 
and line-strength indices. She found that  
 the variations were only visible  in the colours and molecular indices,
more affected by dust, but not in the atomic indices, insensitive to its
presence (MacArthur 2005).

An exception to this, ie, differences in the stellar populations properties of 
KDC and the host galaxy, happens when the KDC is small ($\sim$0.1-0.3 kpc) and it is
in the center of low-massive early-type galaxies (with central velocity
dispersion $\sigma$ around 100 kms$^{-1}$). In these cases the nucleus appear to
be younger and more metal rich than the rest of the galaxy (McDermid et al.
2006). These young nuclei are found, exclusively, in fast rotating galaxies
(see Emsellem et al. 2004 for the definition) and are much less massive than the
'classical' KDC. The formation of these nuclei are probably related to the
accretion of a gaseous small galaxy (second of the scenarios listed above).

Here we present the kinematics and the stellar population of NGC~1700 to
show that it represents an interesting and unique case of KDC. The layout is as
follows: Sec~\ref{sec:props} describes the relevant properties of this galaxy;
Sec~\ref{sec:obs} describes the observations, data reduction and kinematic
measurements;  Sec~\ref{sec:stepop} describes the stellar population study.
Finally, in  Sec~\ref{sec:disc} we present a discussion and our conclusions.

\section[]{Relevant galaxy properties}
\label{sec:props}

NGC~1700 is an elliptical galaxy with a weakly counterotating core
(Franx et al.\ 1989b; Saha \& Williams 1994; Bender et al. 1994; Statler et al.
1999). The galaxy is  fairly bright, with M$_{B}=-22.3$, and massive, with
$\sigma=230$ kms$^{-1}$ (Bender et al. 1992). Its rather high surface brightness
and uncharacteristically small effective radius (r$_{\rm eff} \sim 14"$ (2.6
h$^{-1}$/kpc) (Franx et al.\ 1989a), put it slightly
less than 2 $\sigma$ off the FP in the sense of having unusually low M/L for
galaxies of this mass. The galaxy is boxy inside 3~arcsec and discy outside this
radius (Carollo et al. 1997).  The ellipticity continuously rises outward and
shows structure clearly identified with the variations in the fourth order
cosine term. The discy isophotes and the peak in ellipticity indicate the
presence of a stellar disc between 3 and 10~arcsec. 
NGC~1700 shows several characteristics of being the result of a
merger  that happened  $\sim$5.5-8.3 Gyr ago (Schweizer \& Seitzer 1992).
However, Statler et al. (1996) claimed that the radially increasing prograde
rotation in the main body of the galaxy implies that the major merger event was
not responsible for the counterotating core. They suggest that the KDC in
NGC~1700 is more likely  the result of a merger of 3 or more stellar systems
2-4  h$^{-1}$ Gyr ago, based on N-body simulations (Weil 1995). The surface
brightness profile of this galaxy is well fitted with a Sersic law with $n=5.5$
(Trujillo et al\ 2004) and does not show any $"$break$"$ or slope change at the
position of the KDC (Brown et al.\ 2000).

\section[]{Observations, Kinematics and absorption line-strengths}
\label{sec:obs}

Long-slit spectroscopy was acquired for this galaxy with the blue channel of the
ISIS spectrograph on the 4.2m William Herschel Telescope on 20-22 Nov 2000
within a sample of 9 elliptical galaxies covering a range of masses  and in
isolated environments (see Bergmann 2002).  This configuration gives a
wavelength coverage from 3650 to 5650~\AA~at resolution 2.63~\AA (FWHM).  Four
exposures of $\sim$35m, with the slit positioned along the major axis, were
taken for this galaxy in two nights.
These observations were interspersed with flat-field and arc-lamp
spectra. A selection of F,G, and K stars from the MILES library
(S\'anchez-Bl\'azquez  et al.\ 2006a; Cenarro et al.\ 2007) were also observed
to be used as kinematical templates and for calibration purposes.  Standard
data reduction procedures (flat-fielding, cosmic-ray removal, wavelength
calibration, sky subtraction and fluxing) were performed with {\tt reduceme},
which allows a parallel treatment of data and error frames and
provided an associated error file for each individual data spectrum.

\begin{figure}
\resizebox{0.5\textwidth}{!}{\includegraphics[angle=-90]{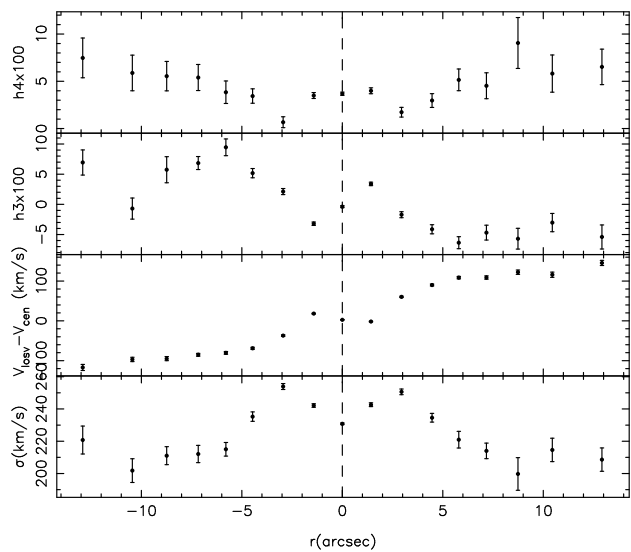}}
\caption{High order Gauss-Hermite moments, radial velocity and velocity
dispersion versus  the radius of NGC~1700.\label{fig:kinematics}}
\end{figure}

Kinematical measurements were performed with the penalized pixel
fitting method ({\tt ppxf}, Cappellari \& Emsellem 2004). The first
two moments, line-of-sight velocity (LOSV) and velocity dispersion (LOSVD) were
derivet and also the higher 
order Gauss-Hermite moments, h3 and h4, which measure deviations of the LOSVD
from a pure Gaussian. Figure \ref{fig:kinematics} shows the results for
NGC~1700.

\begin{figure}
\resizebox{0.5\textwidth}{!}{\includegraphics[angle=-90]{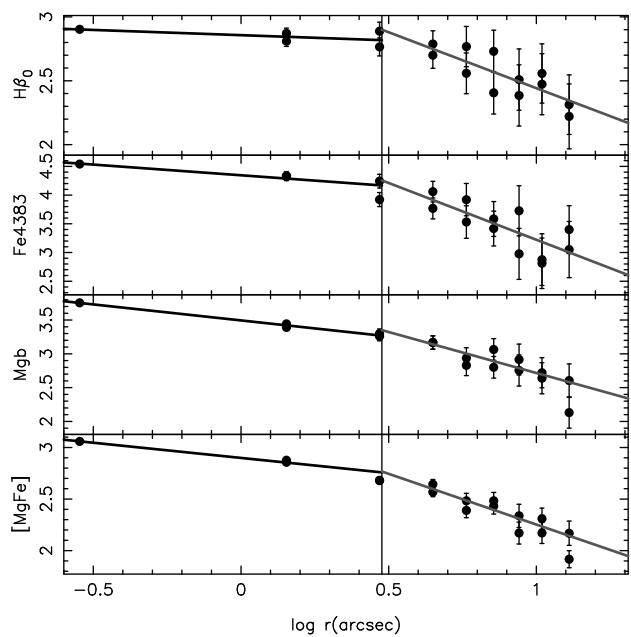}}
\caption{Line-strength indices along the radius for the galaxy NGC~1700. Black and grey
lines show a linear-fit for the regions inside and outside the KDC respectively. The binning
have into account the seeing of the observations (FWHM$\sim 1.3 "$).
\label{fig:line-strength}}
\end{figure}

Line-strength indices with the Lick definition (Trager et al.\ 1998) and the
metallicity-insensitive  index H$\beta_o$ (Cervantes \& Vazdekis 2009) were
measured in our spectra previously degraded to a total  width\footnote{total
width = $\sqrt{(FWHM_{\rm inst}^2+FWHM_{\rm gal})}$, where FWHM$_{\rm inst}$ is
the instrumental resolution and FWHM$_{\rm gal}$ the Doppler widening due to the
random motion of the stars.} of 14~\AA~(FWHM) to match the LIS-14~\AA~ system
proposed in Vazdekis et al.\ (2010). This avoids us to make any correction  for
the velocity dispersion of the galaxy, which can introduce systematic trends in
the relation of the  line-strength indices with other quantities.
Figure~\ref{fig:line-strength} shows the variation of some selected
indices with radius for NGC~1700. A linear fit, weighting with the index errors
and discarding the seeing-affected central 1.3~arcsec region, is shown inside
and outside the radius of the KDC. 

\section{Stellar populations}
\label{sec:stepop}

\begin{figure}
\resizebox{0.6\textwidth}{!}{\includegraphics[angle=-90]{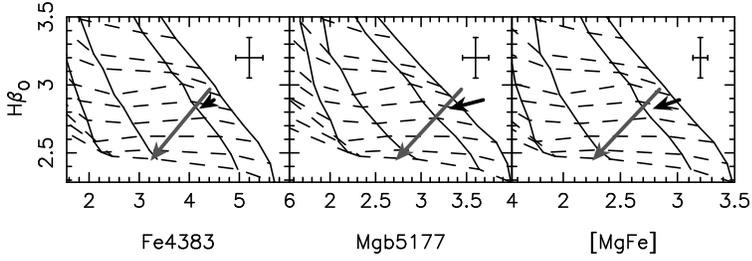}}
\caption{Index-index diagrams comparing the age-sensitive H$\beta_o$ index (Cervantes \& Vazdekis 2009)
with 
different metallicity-sensitive indices in the LIS-14\AA system (Vazdekis et al. 2010).  
Over-plotted are the predictions  of
Vazdekis et al.\ (2010) models of constant age (dashed lines) and metallicity
(solid lines). The values of the different represented ages are (from top to
bottom:  1.8, 2.2, 2.8, 3.5, 4.5, 5.6, 7.1, 8.9, 11.2, 14.1, 17.8 Gyr, while the
values of metallicities are (from left to right: [Z/H]=$-0.4$, 0.0, +0.22. The
linear fits from Fig.~\ref{fig:line-strength} are represented in the Figure. The
region inside the KDC is plotted in black while we represent, in grey,
the region outside the KDC.  The arrow indicates the direction toward
larger radius in each case. Typical errors on each index are plotted as error bars 
on the right upper corner of each panel.
\label{index-index}}
\end{figure}

Figure~\ref{index-index} shows some index-index diagrams comparing different
metallicity-sensitive indices with the optimized age-sensitive index H$\beta_o$.
Over-plotted are the predictions of Vazdekis et al.\ (2010)
single-stellar-population (SSP) models for different ages and metallicities. 
Note that H$\beta_o$ provides a virtually orthogonal model grid. Over-plotted
are the linear fits inside and outside the KDC that are shown in
Fig.~\ref{fig:line-strength}. 

We have derived ages and metallicities by linearly interpolating the models to
the values of the indices in the [MgFe]$'$ vs. H$\beta_o$ diagram. The [MgFe]$'$
index is insensitive to deviations from [Mg/Fe]=0, which occurs in giant
ellipticals and which bias the  ages, depending on the metallicity
indicator in use (Thomas, Maraston \& Bender 2003). To measure these deviations,
we also have obtained metallicities using other indicators which are more
sensitive to variations of Fe (Fe4383) and Mg (Mgb).  The ages inferred from the
different panels are similar, but there are differences in the metallicity
that can be used to estimate the abundance ratios; if a galaxy spectrum is
enhanced in Mg/Fe, we obtain  a higher metallicity when Mgb is used instead of
Fe4383. The metallicity difference obtained in this way,  [Z$_{\rm Mg}$/Z$_{\rm
Fe}$], is a good proxy for the abundance ratio [Mg/Fe]. In fact, a linear relation
between the ratio Z$_{Mg}$/Z$_{Fe}$ and the  [Mg/Fe] values derived directly from 
the Thomas et al.\ (2003) models -- which specifically take into
account non-solar abundance ratios-- was found  in Vazdekis et al.\ (2010) (see also 
S\'anchez-Bl\'azquez et al. 2006b; de la Rosa et al. 2007).

\begin{figure}
\resizebox{0.5\textwidth}{!}{\includegraphics[angle=-90]{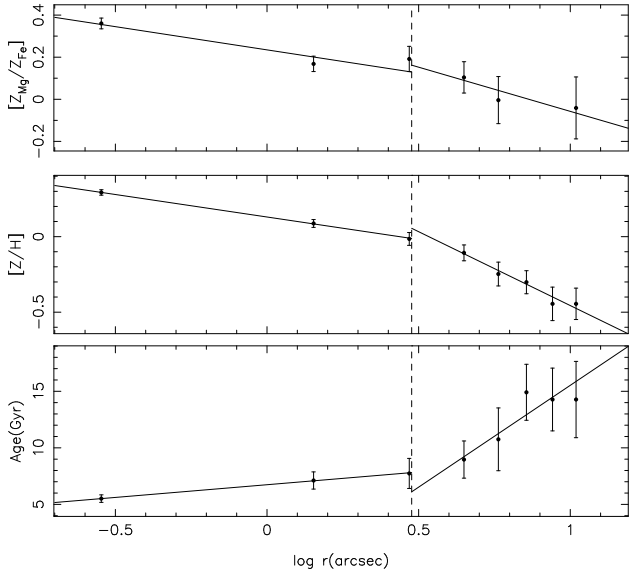}}
\caption{Stellar population parameters of NGC~1700 as a function of radius.
The area corresponding to the KDC is indicated with a vertical dashed line.
To guide the eye, linears fits inside and outside the KDC are also plotted.
\label{fig:ssp_gradients}}
\end{figure}

Figure.~\ref{fig:ssp_gradients} shows the variation of the mean\footnote{Note that 
these values are SSP-equivalent values, and they would be biased towards the 
most luminous stellar populations} age, metallicity and
[Z$_{\rm Mg}$/Z$_{\rm Fe}$] with radius for NGC~1700. Unlike the rest of the galaxy,
 the KDC region shows a clearly younger
and nearly constant SSP-equivalent age of $\sim$6 Gyr. 
The KDC also shows a
metallicity and [Z$_{\rm Mg}$/Z$_{\rm Fe}$]  gradient that is flatter than in the main body of the galaxy.

\begin{figure}
\resizebox{0.3\textwidth}{!}{\includegraphics[angle=-90]{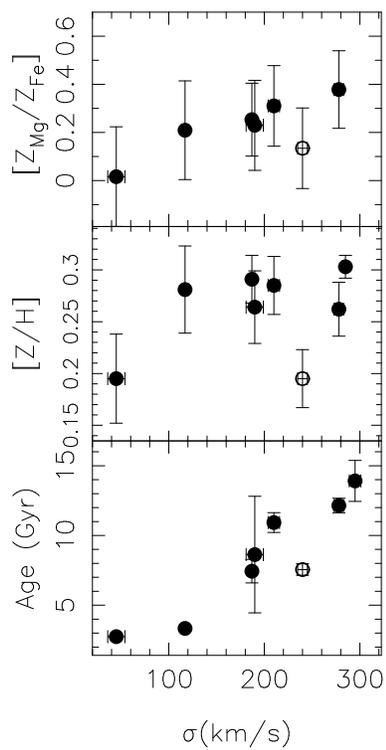}}
\caption{Age, metallicity (measured with a [MgFe]-H$\beta_o$ diagram), and  the
abundance ratio proxy Z$_{\rm Mg}$/Z$_{\rm Fe}$ measured in the central spectra
(r$_{e}$/10) of a sample of field ellipticals of varying mass is plotted vs.
$\sigma$. The open symbol represents NGC~1700.
\label{fig:central_values}}
\end{figure}

Due to the presence of gradients in elliptical galaxies, and to the fact 
that these do not correlate clearly with any other general property, 
if we want to know the stellar populations of the core in relation 
to the main body of the galaxy, we need to 
compare the central stellar population parameters of NGC~1700 with those 
of other galaxies when plotted against $\sigma$ central.
It is well known that the stellar populations in the elliptical family are very
correlated with this parameter. For this purpose we have extracted the central
spectra inside an aperture of $r_e/10$, where $r_e$ represents the effective
radius of the galaxy, for the complete sample of this observing run, consisting
of isolated ellipticals covering a range in mass. This radius, in the case 
of NGC~1700 corresponds approximately to half of the radius of the KDC. We correct for the effect of
rotation  before extracting the central spectra. We also correct for the
presence of emission lines in two galaxies NGC~1172 and NGC~661 using {\tt
gandalf} (Sarzi et al. 2006). We have measured the indices and $\sigma$ and
calculated the metallicities with different indicators as above.
Figure~\ref{fig:central_values} shows the age, metallicity and  $Z_{Mg}/Z_{Fe}$
in the central $r_{\rm eff}/10$ for our sample of galaxies as a function of the
central $\sigma$. It can be seen that NGC~1700 is younger and have  a lower
[Z/H] and $[Z_{Mg}/Z_{Fe}]$ than what it would be expected, giving its $\sigma$,
from the relation defined by the rest of the sample.

\section{Discussion}
\label{sec:disc}

We have shown that the KDC in NGC~1700 have a distinct stellar population than
the rest of the galaxy. This is probably the first time that such feature has
been found in a 'classical' KDC. Although claimed by previous authors (see the
introduction), this was, in most cases, due to the presence of a dusty disc,
affecting colours and molecular indices, or an artifact of plotting the profiles
in linear instead of in logarithmic scale, while the line-strength profiles
follow a power-law. McDermid et al.\ (2006) already found younger components in
the KDCs of some early-type galaxies, however there are important differences as
(1) the size of the young nuclei in the McDermid et al. (2006) sample is never
larger than 0.3 kpc, while the KDC in NGC~1700 is $\sim$0.8 kpc and (2) all the
galaxies with younger nuclei in McDermid et al. (2006) have $\sigma$$\sim$100
kms$^{-1}$, while NGC~1700 is massive, with $\sigma$$\sim$230 kms$^{-1}$.

As described in the introduction, four  scenarios have been proposed to
explain the formation of these structures: 
(1) geometrical projection of stellar orbits in the core of a triaxial system; 
(2) a central disc formed from accreted gas with an initial counter-rotating
orbit; \footnote{Hau \& Thomson (1994)  also proposed that  the spinning up of the halo of a
previously non-rotating elliptical that contains a small, high surface brightness disc by 
a flyby encounter could produce kinematically decoupled structures. The elliptical is left with a central inner disc rotating in a different
direction than the halo. } 
(3) a central disc formed during the merger of two spirals and 
(4) accretion of a stellar companion initially counter-rotating with respect to
the galaxy.
In addition to these  four scenarios, we also considered the possibility
that  the KDC was the end-of-view of a bar that within a hot disc (this 
galaxy has a stellar disc from 3 to 10 arcsec).
 Such system would explain the counter-rotating feature
in the center and the decrease in $\sigma$ observed in that region due to the
ordered motion of the bar (e.g., Bureau \& Athanassoula 2005; de Lorenzo-C\'aceres 2008). However we do not
see any hint of counter-rotation in the stellar disc of this galaxy, as it would
be expected if the bar form from it. Only in the
case that the disc contribution to the total light in the region we are sampling is negligible,
this possibility cannot be fully discarded.

Scenario 1 does not seem to apply to NGC~1700 as the stellar populations should not 
differ  in the transition to the KDC. Such change is, however,
possible for scenarios 2, 3 and 4. 

Scenario 2, as we already mentioned,  is most likely to form the small KDCs
found in McDermid et al. (2006), as the involved mass would be compatible with
that contributed by an irregular gaseous galaxy\footnote{Using stellar population 
models we estimate a stellar mass for the KDC of 8.9$\times$10$^{10} M_{\sun}$}.

Regarding the third scenario, NGC~1700 shows several signatures of having
experienced a merger with a gaseous system (such as tidal tails, see eg., Brown
et al. 2000). Ages of this merger have been estimated using fine structure ages
(Schweizer \& Seitzer 1992; ~ 6.0$\pm$2.3 Gyr and  dynamical ages (Statler 1996;
2.7=5.3 Gyr). These ages are consistent with our age estimate for the KDC of
this galaxy. The central value of the abundance ratio, lower than expected for 
its $\sigma$, is also compatible with
our estimate, as we would expect [Mg/Fe] to be lower if the gas comes from
processed gas in spiral galaxies (Thomas 1999). 

 However, kinematical arguments, as the low rotation and low  skewness h3
 would not be expected if the central component was produced
in a gaseous merger (e.g., Bak 2000; Naab et al. 2006), ie., are not expected
if the central kinematical component was a disc as in the case of, for 
example, IC~1459.
Furthermore,  Statler et al.\ (1996) already pointed out that 
the radially increasing prograde  motion with radius
at R$>$50 arcsec argues against a single merger event in this galaxy.
A single merger would not deposit prograde
orbiting stars at large radius and retrograde at small radius. NGC~1700 must
have merged with or ingested at least two other stellar systems to account for
its present dynamical structure.

Scenario 4 has been modelled by Balcells \& Quinn (1990). Balcells (1991) obtained
coefficients of skewness $\sim$0.2 corresponding to h3$\sim$0.03, very near to the
values we measure in the core of NGC~1700  $\sim 0.05$. Furthermore, Balcells \& Quinn (1990)
confirm Kormendy's (1984) conjecture that the central velocity dispersion of the
remnant should be slightly depressed, as the light here is contributed
significantly by the core of a less massive system. We also observe a
depression of $\sigma$ in the central region of NGC~1700 (note that 
this feature is not observed in IC~1459 where the kinematically decoupled
structure is more likely produced by a disc). In the scenario  proposed by Balcells \& Quinn (1990) we
would expect to find lower age, metallicity and [Mg/Fe] as expected for a low
mass early-type galaxy. All this holds for the center of NGC~1700.

We conclude that the most likely scenario for NGC~1700 is the ingestion of a
small stellar companion on a retrograde orbit as the cause of counterotating
core, with a later gaseous merger, responsible for the tidal tails observed in
this galaxy.

\section*{Acknowledgments}
It is a pleasure to thank Adriana de Lorenzo-C\'aceres, Jes\'us Falc\'on-Barroso,
Inma Mart\'{\i}nez-Valpuesta \& Ignacio  Trujillo for useful and stimulating 
conversations.
PSB is supported by the Spanish Ministerio de Ciencia e Innovaci\'on (MICINN)
through the Ram\'on y Cajal programme and by the European Union through its
FP7 program.
Based on observations made with the WHT operated on the
island of la Palma by the ING in
the Spanish Observatorio del Roque de los Muchachos of the Instituto
de Astrof\'{\i}sica de Canarias. This research has been supported by
the Spanish Ministry of Science and Innovation (MICINN) under the
grants AYA2007-67752-C03-01 and AYA2007-67752-C03-02

\label{lastpage}
\end{document}